\documentclass[12pt,aps]{revtex4}
\usepackage{epsfig}
\usepackage{amsmath,amssymb,color}
\usepackage{epstopdf}
\parskip=\medskipamount

\newcommand{\Tr}{\mathop{\rm Tr}\nolimits}
\newcommand{\beq}{\begin{equation}}
\newcommand{\eeq}{\end{equation}}
\newcommand{\be}{\begin{equation}}
\newcommand{\ee}{\end{equation}}

\begin{document}

\title{Flavored extended instanton in QCD}

\author{A. Gorsky$^{1,2}$, A. Grekov$^{3,4}$}

\affiliation{
$^1$ Institute for Information Transmission Problems of the Russian Academy of Sciences, Moscow,
Russia, \\
$^2$ Moscow Institute of Physics and Technology, Dolgoprudny 141700, Russia \\
$^3$ Physics Department , Stony Brook University, USA\\
$^4$ Steklov Mathematical Institute of Russian Academy of Sciences, Gubkina str. 8, 119991, Moscow, Russia
}

\begin{abstract}
In this paper we discuss new flavored space-like defects in confined QCD
which can be considered as the Euclidean extended instantons carrying the 
topologically quantized currents. 
We focus  on the simplest  1d space-like defect - the S-Skyrmion
solution extended in one space coordinate and localized in Euclidean time. 
It can be identified both in the
holographic QCD and in the Chiral Perturbation Theory(ChPT).
The Skyrmion charges get transformed into the corresponding currents for S-Skyrmion.  
The analogy with the Thouless pump and the quantum phase slip  phenomena is mentioned.

\end{abstract}
\maketitle
\section{Introduction} 

The space-like branes or S-branes have been suggested in \cite{sg} in  
field theory and in the string theory framework. These objects  were
identified as  defects of different codimensions 
localized in time. The conventional instanton localized in the Euclidean time is the simplest example. 
The interpretation of  the extended space-like defects
is a subtle issue and two scenarios have been suggested in \cite{sen}. Both of them involves
the brane configurations with tachyons - the unstable D-brane or the $\bar{Dp}- Dp$ state.
The potential for the tachyon field allows  the kink-like solution, and the space-like brane
is interpreted as a tachyon kink  in the time direction. It describes
some process in  field or string theory, in particular it was suggested to describe the
decay of unstable branes into strings \cite{sg2,gaiotto,lambert}. The precise example of the
creation of the time-like objects from the S-branes has been found in \cite{hashimoto2003}.

We shall focus on the similar objects in the conventional  QCD in confined phase where baryon 
is identified as the Skyrmion \cite{witten}. In the ChPT it is  solution to the equation of motion with the baryonic charge 
which provides its topological stability. Upon the proper identification of the electromagnetic 
current \cite{witten2} it can be shown that it can  carry an electric charge as well
and enjoys the fermionic statistics \cite{witten3} due to the 5d Chern-Simons term for the
flavor group. The stabilization of a size is provided by the Skyrme term. The solution can be obtained from the holonomy
of the instanton solution if one additional artificial dimension is added \cite{atiyah}. This old observation looked a bit puzzling 
for a while but it gets the clear-cut
explanation in the 5d holographic QCD \cite{tong}. Baryon aka Skyrmion is nothing but the 
instanton solution in the holographic QCD on the worldvolume of flavor $N_f$ D8 branes \cite{son}
which is localized in three space coordinates and radial holographic coordinate $r$. 
Hence it is extended in time direction and can be considered as a particle. The instanton charge gets identified with the baryonic charge.
If the chiral condensate is taken into account for the Skyrmion solution the
dyonic instanton in the holographic QCD should be considered \cite{gk}. 
The dyonic instanton solution involves the nontrivial space profile for the
tachyonic scalar in the bi-fundamental representation. Baryonic mass
becomes the chiral condensate dependent \cite{gk} and the partial restoration
of the chiral symmetry in the core of the solution takes place .

More recently the different picture for the high-spin baryon was suggested \cite{komargodski}.
The  baryon  for $N_f=1$  can be thought of as the finite $\eta'$ 
domain wall with the chiral boundary excitations very much as the quantum Hall droplet. 
The baryon with high spin was identified as the chiral excitation at the boundary  and the baryon charge
was related to the 2-form symmetry current.

In this paper we consider new topologically nontrivial solutions in 5d holographic QCD  which 
are flavor instantons localized in Euclidean time therefore being the examples of space-like defects
or "`extended instantons"'. Our 
main example will be the space-like S-Skyrmion that is the defect with 
one-dimensional worldsheet extended for example in $x_3$ coordinate. 
Its topological charge describes the map of the 
asymptotics of $(r,t_E,x_1,x_2)$ 4-dimensional space into the flavor gauge group.
Contrary to the Skyrmion when the topological charge yields the baryonic charge density in
this case the space component of the quantized topological current is generated. Hence we obtain a kind of
instant one-dimensional defect carrying the topological quantized  current  in the conventional 
confined QCD. 
We question if S-Skyrmion enjoys the counterparts of the phenomena familiar 
for the Skyrmion: electric and axial charges,
fermionic statistics,
induced decay in the monopole background.
We shall argue that the S-Skyrmion hosts the electric and axial currents. The fermionic statistics 
of the Skyrmions gets mapped into the specific properties of the
S-Skyrmion as well. The possible analogue of the Callan-Rubakov effect
is suggested.

The topologically quantized currents we consider have some analogy
with the Thouless pump phenomenon \cite{thouless} 
which concerns the topologically quantized current at the
interval. The quantization is supported by the peculiar
topological invariant which involves the integration over the period
of the drive. Physically if there is periodic drive applied to the
insulator of the finite size the charges of the opposite sign at the insulator
boundaries are generated.
Recently the related effect, when  the topologically
quantized work occurs, has been found \cite{moore,sasha2}. 
In this case the quantized work of the opposite sign  has been performed 
at the boundaries of the system with the gapped bulk.

One more example of the similar nature is the quantum phase slip phenomenon
in 1+1 dimensions  \cite{flip}. It is to some extend dual to the superconducting
current. There is the instantaneous topological current of
vortices across the superconducting wire. Microscopically the
amplitude of the Cooper condensate vanishes locally and the the phase
rapidly rotates at $2\pi k$. This flip blocks the superconducting current.  
In our situation we have the tunneling topological current 
"`across the sample"' as  well. In the dyonic instanton realization
of the S-Skyrmion the analogy is very close - in our case we have 
the small region with the restoration of the chiral symmetry which
supports a tunneling of the topological charge.
However if we realize the S-Skyrmion as pure instanton there is 
no need to have the "`gapless channel"' through the bulk.

In lattice QCD these configurations should provide the finite 
contribution into the partition function since the periodic 
boundary conditions are imposed at the Euclidean space coordinates. 
In fact the non-perturbative  1d defects 
with the peculiar properties have been found in lattice QCD long time
ago (see \cite{zakharov} for the review). They have been named
as percolating monopoles however it is unclear if these configurations
are related to the solutions which we shall describe in this paper.

The paper is organized as follows. First we remind the realization of Skyrmion as
the instanton in holographic QCD. In Section 3 we consider the S-Skyrmion
in holographic QCD and in ChPT and discuss its properties.
Section 4 is devoted to the interpretation of the array of S-Skyrmions
while open questions are formulated in Conclusion.

\section{Skyrmion as flavor instanton}

Let us recall the instanton realization of the baryon in the holographic QCD.
In the Witten-Sakai-Sugimoto(WSS) \cite{witten98, ss1} model at $T=0$ the holographic background 
looks as the cigar-like geometry involving coordinates $(r,\phi)$ supplemented with sphere $S^4$ and
four-dimensional Minkowski space-time. 
The flavor degrees of freedom
are introduced by adding $N_f$ $ D8-\bar{D8}$ branes extended along
all coordinates but $\phi$. The theory on the flavor D8 branes 
upon the dimensional reduction on $S^4$ yields the 5-dimensional Yang-Mills
theory with $SU(N_f)_R\times SU(N_f)_L$ gauge group supplemented with
the Chern-Simons term. The action reads as 
\beq
S= \sigma\int d^4x dz ( h(z) \Tr F_{\mu\nu}^2 +g(z) \Tr F_{\mu z}^2) + S_{CS}
\eeq
where $\mu,\nu = 1,2,3,4$ the metric factors are
\beq
h(z)=(1+z^2)^{1/3} \qquad g(z)=(1+z^2)
\eeq
and $\sigma$ is expressed through the t'Hooft coupling $\lambda$
as $\sigma= \frac{\lambda N_c}{216\pi^3}$.
It yields the Chiral Lagrangian in the 
conventional low-energy QCD and  reasonable values of the
low-energy parameters \cite{ss1}. 

The baryon in the WSS model is identified as the D4 brane wrapped 
around $S^4$ and extended in the time direction. In terms of the 
5d YM theory with the flavor gauge group the baryon is the instanton solution localized 
in $(z,x_1,x_2,x_3)$
coordinates. Consider for example $N_f=2$ case and separate the $U(2)$ flavor gauge
field on D8 branes into the $SU(2)$ field $A(x,z)$ and U(1) field $B(x,z)$.
The solution for the instanton sitting around $(x=0,z=0)$ reads as 
\beq
A_{\mu}= -if(\eta)g_{inst}(x,z)\partial_{\mu}g_{inst} \qquad A_0(x,z)=0 \qquad f(\eta)= \frac{\rho^2}{\eta^2 +\rho^2}
\eeq
where 
$$g_{inst}=  \frac{(z-z_0) -    i(\vec{x} - \vec{x_0})\vec{\tau}}{\sqrt{(z-z_0)^2 + |\vec{x} - \vec{x_0}|^2}} $$
\beq
B_i(x,z)=0 \qquad B_0(x,z)= -\frac{1}{8\pi^2 \lambda \eta^2} [ 1 - \frac{\rho^4}{(\eta^2 +\rho^2)^2}]
\eeq
This solution is nothing but the Skyrmion solution 
and it realizes 
old Atiyah-Manton interpretation  \cite{atiyah}.
The BPST instanton  can be used  as a good approximation since it was argued
in \cite{more,pomarol} that the solution is mainly localized around $z=0$
where the wrap factor can be neglected. 
The radius of the instanton solution in $(x,z)$ space is fixed at the extremum
of the corresponding potential
\beq \label{potential}
U(\rho)\propto \frac{\rho^2}{6} + \frac{1}{320\pi^2 a^2\rho^2}   \qquad \rho_{inst} = \frac{1}{8\pi^2\lambda}\sqrt{6/5}
\eeq
The second term in the potential comes from the Coulomb interaction due to CS term.
The 5d CS term 
provides its fermionic statistics \cite{witten2}. 
The baryonic charge B gets identified as 
\beq
B=\int d^3 xdr( \Tr F_{L}\tilde{F}_{L} - \Tr F_{R}\tilde{F}_{ R}) 
\eeq
where $r$ is proportional to $z$.

The Skyrmion solution can be derived in the conventional ChPT with Lagrangian
\beq
L=-\frac{F_{\pi}^2}{16} Tr(U\partial_{\nu}U^{\dagger})^2 + \frac{1}{32c^2} 
Tr([U\partial_{\mu} U^{\dagger},U\partial_{\nu}][ U^{\dagger}U\partial_{\mu} U^{\dagger}U\partial_{\nu} U^{\dagger}])
\eeq
where the second Skyrme term provides the stabilization of the Skyrmion.
The baryonic charge in ChPT is expressed in terms of the unitary matrix field $U(x,t)$
build from the Goldstone pions 
\beq
B=\int d^3 x \Tr (U^{-1}dU)^3 \qquad U(x,t) =\exp(\frac{i\pi^a t^a}{F_{\pi}})
\eeq
Its topological nature is supported by the nontrivial $\pi_3(SU(2))=Z$
since  asymptotic condition $U(x_1,x_2,x_3 \rightarrow \infty) \rightarrow 1$
provides the
mapping of $S_3$ into the diagonal flavor group.
The solution
has the electric charge due to  $N_c$ fundamental strings attached
to the baryonic vertex or D4 brane \cite{wittenvertex}. The mass of
the Skyrmion is $M\propto \frac{F_{\pi}}{c} $ while its radius is $r_s= \frac{1}{F_{\pi}c}$.
The Skyrmion enjoys the axial and tensor charges as well.

It is possible also to include the chiral symmetry breaking
condensate into the 5d action explicitly via the  
boundary behavior of the additional tachyonic scalar field X in the
bi-fundamental representation. In this case the baryon becomes
the dyonic instanton solution with two quantum numbers
and the tachyon field has nontrivial kink-like 
profile in the space. The mass 
of the dyonic instanton in some regime 
is determined by the chiral condensate \cite{gk}. 
One could say that in the dyonic instanton representation
of the Skyrmion the chiral symmetry breaking is partially
restored at its core.

\section{S-Skyrmion  in holographic  QCD and ChPT}
\subsection{Currents}
Turn now to the space-like S-Skyrmion solution in QCD. First, perform 
the Wick rotation and consider $R^4$ instead of the Minkowski space.
The D4 brane representing the S-Skyrmion  is wrapped around $S^4$ and extended in $x_3$ 
coordinate. Since D4 brane share all coordinates with D8 branes
it amounts to the instanton-like solution in the flavor gauge theory
as for any $Dp-D(p+4)$ system. However this 
instanton has the different interpretation in comparison
with the baryon since it is localized in the Euclidean time
and is extended in one space dimension.

Consider the BPST  solution in the 5d flavor YM theory 
localized in  $(t_E,r,x_1,x_2)$.  It looks the same as the
standard Skyrmion-instanton however is extended along  say $x_3$ 
coordinate instead of  time coordinate. 
\beq
A=A_{inst}(t_E,x_1,x_2,z)  \qquad B=B_{inst}(t_E,x_1,x_2,z)
\eeq
The potential for the size of  extended instanton  is the same as before. 
Therefore similar to the conventional Skyrmion using the approximate
rotational symmetry of the solution we can estimate the energy density of 
the S-Skyrmion as the mass of the conventional Skyrmion $T= M_{Sk} $ and its size 
in the Euclidean time direction is identified with the radius of the Skyrmion
$\delta t_E = \rho_{Sk}$. 

There is
topologically conserved current in 5D 
\beq
J^{5d}=*\Tr F_A\wedge F_A,  \qquad F_A= F_L -F_R
\eeq
which yields the baryonic charge for the Skyrmion.
S-Skyrmion  carries non-vanishing current component 
$J_3$, say along $x_3$ space coordinate along which it is extended 
\beq
J_3= \int dx_1 dx_2 drdt_E \Tr F\tilde{F} = \int_{x_3=const} \Tr F_A \wedge F_A
\eeq

In the ChPT the extended instanton solution saturates the topological charge
representing $\pi_3(SU(2))$. Upon the imposing the asymptotic behavior
in the Euclidean space-time 
$U(x_1,x_2,t_E \rightarrow \infty) \rightarrow 1$  it
maps the three-dimensional sphere into the flavor group.
The topological current density substituting the baryonic charge 
in this case  reads as
\beq
\tilde{J}_3 = \int dt_E d^2x \epsilon_{t12} \Tr (U^{-1}d_t U)(U^{-1}d_{x_1}U)(U^{-1}d_{x_2}U)
\eeq
and involves the integration over Euclidean time and two space-like coordinates $x_1,x_2$.
This means that solution has nontrivial topological current density
only for the time-dependent pion fields.

The next question concerns the electric charge of the solution.
The following term  in the ChPT
is relevant for the electromagnetic current evaluated
at our solution
\beq
L_{wzw} = -\frac{N_ctrQ}{32 \pi^2} 
\epsilon^{\mu\nu\alpha\beta} 
Tr(U^{-1}\partial_{\mu}UU^{-1}\partial_{\alpha}UU^{-1}\partial_{\beta}U)A_{\nu}
\eeq
We can extract this contribution also from 5d action
involving the $U(1)$ connection as well and get the term in the 4d action
\beq
S_{int}= N_c\int dx_3 A^{U(1)}_3 J_3
\eeq
which means that the  S-Skyrmion is coupled to the abelian gauge
potential. The $N_c$ factor in the standard Skyrmion  tells that the $N_c$ fundamental strings
are attached to the baryonic vertex supporting its composite nature 
as state build from $N_c$ quarks with fractional baryonic charge.
Similarly we could assume that S-Skyrmion has the composite nature
and can be thought as the baryonic vertex with $N_c$ fundamental
strings each of them carry the fractional topologically 
quantized current. The $N_c$
fundamental strings attached to the vertex have $(x_3,r)$ worldsheet coordinates.

It is useful to compare this current with the electric current 
in the external magnetic field via CME \cite{cme}. In the hadronic
phase the CME induced electric current has been discussed in \cite{mameda}
It needs for the chiral disbalance induced
by the chiral chemical potential or the time dependent pseudoscalar
field and external magnetic field.
In our study we have the electric current induced by the 
time-dependent pion classical configuration in the Euclidean space
without the external magnetic field.
Remind that the solution is topologically non-trivial due
to $\pi_3(SU(2))=\mathbb{Z}$, hence we can integrate the current density
and obtain non-vanishing $J_3$ current during the whole process. It
is proportional to the topological invariant therefore 
is quantized and do not vanish for our extended instanton solution. 

The Skyrmion carries the axial charge which can be easily 
holographycally seen as
follows. If we treat the radial coordinate 
as time the canonical momentum for the axial gauge  field gets modified 
and reads as
\beq
\Pi_{\mu}^A= E_{\mu}^A + N_c K_{\mu}^A
\eeq
where the second term follows from the CS term. Now consider the Gauss law
constraint in the axial channel $\partial_{\mu}E_{\mu}^A=0$ express 
the electric field in terms of the canonical momentum and impose this
constraint on the Skyrmion state
\beq
(\partial_{\mu}\frac{\delta}{\delta A_{\mu}} - N_c F\wedge F )| Skyrmion>=0
\label{axial}
\eeq 
The variation over the axial gauge field yields the axial current 
at the boundary, hence the Skyrmion which has the non-vanishing
value of the baryonic charge derived upon integration over the
space hence due to (\ref{axial}) has the axial charge as well.
In fact this is the analogue of the Witten effect when the
monopole acquires the electric charge in the presence of the $\theta$-term
due to the modification of the canonical momentum.

We can apply the same logic for the S-Skyrmion which carries the topological
current. The only difference is that we integrate 
the Gauss law constraint over $(x_1,x_2,t_E)$. Hence 
S-Skyrmion carries the axial current as well.

\subsection{S-Skyrmion from the dyonic instanton}

One can consider more general solution for the description of the S-Skyrmion
from the 5d viewpoint. In the conventional Skyrmion case the dyonic instanton
solution has been numerically found in \cite{gk} and takes into account the chiral 
condensate via the boundary condition for the bi-fundamental 
tachyonic scalar $X$. 
\beq
X=\frac{1}{2}(mz +\chi z^3)+ \dots ,\qquad <\bar{q}q>= \frac{N_c}{2\pi}\chi
\eeq
The bifundamental scalar is tachyonic and follows from the mode of the string
connecting flavor $\bar{D}8-D8$ branes. 
The Skyrmion solution in this case involves the nontrivial
profile of the tachyon scalar field in the space 
and it was argued that the chiral
condensate tends to vanish in the Skyrmion core.

We could consider the similar 
dyonic instanton solution for the S-Skyrmion.
The situation starts to remind  
the initial interpretation of the generic  S- brane as the tachyon kink in the time direction
in the $\bar{D}p-Dp$ system \cite{sg}. The
S-brane is assumed to be located at the extremum of the tachyon potential V(T) at $T=0$.
The tachyon field involved into the dyonic instanton solution
has the kink profile in the Euclidean time direction and S-Skyrmion
is located around $X=0$ where the chiral condensate tends to vanish.
The energy density of the
S-Skyrmion will depend on the value of the chiral condensate like
in \cite{gk} for the conventional Skyrmion.

The situation resembles the phase slip phenomenon. Indeed the dyonic
instanton solution provides the instantaneous "`ungapped channel"' in the gapped
bulk very much as the amplitude of the superconducting condensate
vanishes locally allowing the phase to rotate. In the phase slip case the phase of the condensate 
rotates almost instantaneously yielding the non-vanishing 
topological invariant which measures the jump of the phase of the condensate. 
In our case the pions play
the role of the phases of the chiral condensate and the 
topological invariant measures the jump of the pionic phase as well.

The conventional Skyrmion is fermion due to the 5d CS term \cite{witten3}. It is natural to address
the question concerning the counterpart of the fermionic statistics for the S-Skyrmion.
Instead of rotation in $(x_1,x_2,x_3)$ we have very similar topological arguments 
concerning rotation in $(t_E,x_1,x_2)$ in the Euclidean space.  That is it is 
"fermion" in this Euclidean space and the Pauli principle naively forbids the S-Skyrmions
at one point in  $(t_E,x_1,x_2)$ space. However there are some subtleties concerning
analytic continuation into the Minkowski space and we shall discuss the different aspects
of the S-Skyrmion statistics elsewhere.

\subsection{Finite worldline of S-particle}

The action of the S-particle  solution extended in $x_3$  is proportional to the length of $x_3$ coordinate - so for non-compact $x_3$ it is infinite, hence to have the finite action  we need to consider the S-particle with the
finite length. This can be achieved by periodic $x_3$ or by imposing the proper boundary 
condition providing its termination  at the ends of finite interval. 
Consider first the periodic coordinate.  Since the space-like particle could be considered as a baryon with $x_3$ playing the role of euclidean time, to find all the possible classical string configurations we need to classify the classical periodic solutions of equations of motion for the particle moving in the inverted potential (\ref{potential}). 
It is easy to see that the periodicity requirement leaves only solutions, for which the effective instanton radius $\rho$ does not depend on $x_3$ - the particle sitting at the top of the potential. 
However these solutions could still move in the physical space and in the inner sphere $S^3$. 
The action on such solutions is equal to
\begin{equation}
S_{\text{string}} = S_{kin} + L_{x_3} \, \frac{4 \pi}{g_{5d}^2}
\end{equation}
where $L_{x_3} $ is a size of $x_3$ circle.

To discuss the second possibility  and explain the possible termination of the S-Skyrmion 
at some value of space coordinate consider
the analogous process - the decay of the Skyrmion in time. The key point is
that the baryonic current gets modified in the external electromagnetic field
and reads as
\beq
B_{\nu}=\frac{1}{24\pi^2} \epsilon^{\mu\nu\alpha\beta} 
Tr(U^{-1}\partial_{\mu}UU^{-1}\partial_{\alpha}UU^{-1}\partial_{\beta}U) -
\frac{1}{8\pi^2} \epsilon^{\mu\nu\alpha\beta}\partial_{\nu}[(A_{\alpha}
Tr(Q(U^{-1}\partial_{\beta}U + \partial_{\beta}UU^{-1}))]
\eeq
In the monopole background the Bianchi identity for the gauge field 
is violated $\partial_{\nu} \tilde{F}_{\nu\mu} = J^{mon}_{\mu}$  and 
the density of the baryon current is not conserved 
\beq
\partial_{\nu}B_{\nu}\neq 0
\eeq
Integrating  baryon charge over the 3-dimensional space one gets 
the rate of the baryon decay in the monopole background
\beq
\frac{dB}{dt}\propto \rho^{mon}  \partial_{t} \pi^0
\eeq
It is the Skyrmion realization of the Callan-Rubakov effect \cite{rubakov,callan}.
The unwinding of the Skyrmion occurs through the time dependent
pion field. 
The process has been identified in holography as well \cite{hong} where 
the wrapped D4 Skyrmion gets dissolved in the D6 monopole string. 

The S-Skyrmion also undergoes the termination at the S-monopole. To 
identify this effect assume that the monopole current is directed
along $x_3$ coordinate hence $ J^{mon}_{3} \neq 0$. The 
non-conservation of the component of the baryonic current now reads as
\beq
\frac{d B_3}{dx_3} = J^{mon}_3 \partial_{x_3} \pi^0
\eeq
Integrating this equation over $\int dx_1 dx_2 dt_E$ we obtain 
the rate of "`termination"' of the instanton extended along $x_3$ coordinate. 
Since the S-Skyrmion is represented by the same D4 vertex with the
rotated worldsheet the termination of S-brane can occur similarly
via dissociation at the D6 brane represented by the monopole string. 

\subsection{Finite temperature}

Let us consider the case of finite temperature that 
is periodic Euclidean time. Since the Euclidean
time is involved in our S-Skyrmion, solution  gets modified
into the caloron. The caloron configuration \cite{caloron}
for $SU(N)$ gauge group can be thought 
as the composite object involving $N$ constituents 
with magnetic and fractional instanton $1/N$ charges.
The total magnetic charge vanishes while the total
instanton charge equals one. The constituents are 
distributed along the thermal direction at distances dictated 
by the value of the Polyakov loop. 

The caloron is described by the following Nahm equation for the dual gauge field 
\begin{equation}
\frac{d}{dz} \hat{A}_i + [\hat{A}_0,\hat{A}_i] - \frac{1}{2} \epsilon_{ijk} [\hat{A}_j,\hat{A}_k] = i \sum_A \delta(z-\mu_A) \Tr (\sigma_i \text{Im} \overline{\lambda}_A \lambda_A)
\end{equation}
where $\exp(2\pi i \mu_A)$ are the eigenvalues of the Polyakov loop, and each source in the r.h.s corresponds to the magnetic constituent.
In the simplest case the solution reads as
\begin{equation}
A_{\mu}^a = \bar{\eta}_{\mu \nu}^a \, \Pi(x) \partial_\nu \Pi^{-1}(x) 
\end{equation} 
where $\bar{\eta}_{\mu \nu}^a$ - anti-t'Hooft symbol
and $\Pi(x)$ has the form:
\begin{equation}
\Pi(x) = 1 + \frac{\pi \rho^2  T}{r} \frac{\sinh(2\pi r  T)}{\cosh(2\pi r  T) - \cos(2\pi t_E  T)}
\end{equation}
where $T$- is temperature, $\rho$ -is a size of a solution and $r^2 = x_1^2 + x_2^2+ z^2$.

In our case we have such caloron solution in the flavor 
gauge group hence our S-Skyrmion
gets defragmented into the $N_f$ constituents  with additional
"`flavor magnetic"' charges and fractional topological numbers.
Since the topological number now measures the current this means 
that the total current can be represented in the finite
number of components with the fractional current. Since the S-brane 
here is the baryonic
vertex with $N_c$ strings the fractionalization implies
the formation of $N_f$ groups with $\frac{N_c}{N_f}$ strings
in each group.

\subsection{Extended instanton and lattice QCD}

The lattice QCD deals with the Euclidean space-time with
periodic boundary conditions. This set-up is suitable
for the search of the S-Skyrmion with quantized currents.
Indeed our S-Skyrmion  is localized at $T^3\times R^1$ and the
Chern number can be defined in this case. The instanton solution
in this geometry gets fractionalized and can be treated 
as the bound state of the fractional instantons with
fractional topological currents. We expect that such 
1d closed loops extended along one space coordinate
should be observed in lattice QCD studies.

In fact the whole Zoo of defects has been observed
on the lattice (see \cite{zakharov} for the review).
They involve defects with 1d, 2d and 3d worldvolumes.
The 1d defects found in the lattice QCD were interpreted
as the monopoles, moreover two types of monopole
configurations - one of IR nature while the second of UV 
nature were observed. It would be interesting
to compare properties of these observed 1d defects 
with properties of S-Skyrmion. Note that the holographic
classification of  QCD defects can be found in \cite{gzz}.

One more point worth to be mentioned. It is known 
from the lattice QCD that  all eigenfunctions of the
4d Euclidean Dirac operator are delocalized \cite{osborn}
in confined phase while there is the mobility edge in the
deconfined phase
(see \cite{gl} for the recent holographic interpretation).
The low-energy QCD is treated as the random chiral matter
and the pion decay constant $F_{\pi}$ defines the diffusion
coefficient. We could speculate that the extended instantons
of different codimensions
could provide the ungapped channels for the delocalization via a
kind of percolation mechanism however this point certainly
deserves the further study. If it is true it would be a kind
of fracton picture (see \cite{fracton} for review) for the
transport of Dirac operator modes.

\subsection{2d analogue}.
 
Let us comment on the similar baby S-Skyrmion solution in the 2d $N_f=1$  QCD within 
3d $U(1)_L\times U(1)_R$ flavor gauge group in the holographic description 
\cite{zahed} involving $(t,x,r)$ coordinates. 
The Lagrangian of the
model involves the tachyonic bi-fundamental scalar similar to the
4d case.
The conventional
vortex solution to the equation of motion in the gauge theory in 
$(r,x)$ space yields the analogue of the
Skyrmion in 2d propagating in time and having the topological charge
$B_2=\int dx A_x$ well defined if $x\in S_1$. 

Now let us make the
Wick rotation and consider the vortex solution
in the $(r,t_E)$ plane instead. It is localized in Euclidean time
and extended in the $x$ coordinate that it is a kind of extended instanton.
The topological charge of such space-like defect solution in 3d flavor YM gauge theory
supplemented with 3d CS term reads as
\beq
Q_2=\int dt_E dr *F= \delta \int dt_E A_0 - \delta \int dr A_r
\eeq
where we take into account that the integrand is the total derivative.

What is the interpretation of this solution if any? First note that
there is topologically conserved axial 3d current density in the theory
\beq
J^{3d}=*(F_L -F_R)
\eeq
which amounts to the non-vanishing axial  current component 
\beq
J_x=Q_2
\eeq
on our solution.
Hence we have a impulse-like axial current along the space coordinate. 

In the gauge $A_0=0$ the phase of the chiral condensate 
is identified with the $\int dr A_r$ hence our topological
invariant for the baby 1+1  S-Skyrmion is just the jump of the
phase at the instantaneous topological current. The situation
is very similar to the quantum phase slip phenomenon in the
superconducting wire. The only difference is that we have 
substituted the Cooper condensate  by the excitonic condensate.
In some sense this subsection provides the holographic 
realization of the quantum phase flip phenomena.

\subsection{Analogy with the Thouless pump phenomenon}

An interesting analogy with the Thouless pump phenomenon \cite{thouless}
worth to be mentioned. It was argued
in \cite{thouless} that  for the periodic driving process
a pump of a charge at the boundaries of the gapped space
interval can be identified and is topologically protected.
It is to some extend the non-stationary  analogue of the TKNN invariant
defined for the stationary case which yields the Hall conductivity. The invariant expression for the
current reads as \cite{thouless}
\beq
Q=\frac{1}{T} \int ^{T}_{0} dt d^2 k 
Tr (U^{-1}\partial_{t}UU^{-1}\partial_{k_1}UU^{-1}\partial_{k_2}U)
\eeq
and it involves the nontrivial mapping of the 
$(t,k_1,k_2)$ space into the group of rotation of the
ground state of the system. It is a version of the
Chern number for the Berry connection. The corresponding
current is quantized due to its topological nature.

More recently the similar Thouless pump phenomenon has
been found for the topologically quantized work instead of the current flow \cite{moore}.
Moreover the Chern number can be identified not only necessarily 
for the momentum space but for the coordinate space as well \cite{sasha2}.
The corresponding expression for the 
Chern number yielding the topologically quantized work reads as 
\beq
Q=\frac{1}{T} \int ^{T}_{0} dt d^2 x 
Tr (U^{-1}\partial_{t}UU^{-1}\partial_{x_1}UU^{-1}\partial_{x_2}U)
\eeq
The total energy is conserved of course but there is topologically
quantized work done at the edges with the opposite signs.

Our expression for the topologically quantized current is of the
same nature. The only difference is that we find the extended
instanton not in Minkowski but in the Euclidean space hence 
its interpretation is a bit different. However as we have 
mentioned above the tunneling interpretation of the instantaneous 
current is relevant for the quantum phase slip.

In the Thouless pump or energy pump the Berry phase interpretation
involves the matrix of unitary rotation of the state $U(x,t)$. We could question
if the Berry phase interpretation of our invariant is possible.
The matrix $U(x,t)$
involved into the Chern number of the S-Skyrmion indeed can be interpreted as the
chiral rotation of the ground state in the chirally broken
phase. Namely the Chiral Lagrangian can be derived from the quark
fermionic determinant if we assume that the pions provide 
the chiral phase of the quark mass.
Hence from the quark viewpoint  a kind of Berry connection can be defined for
the external time-varying pion field.

\section{Multiple space-like defects}
\subsection{Towards the monopole string}

As it was discussed in  \cite{rossi} the BPS monopole solution could be obtained as an infinite sequence of instantons:
\begin{equation}
\Pi(x,t_E) = \sum_{n= -\infty}^{+\infty} \frac{1}{\beta^2 (r^2 + (t_E - \frac{2 \pi n}{\beta})^2)}
\end{equation}
after a special gauge transformation with the help of the group element:
\begin{equation} \label{rossigauge}
U(x,t_E) = \exp(-i \tau_a \frac{x_a}{|x|} \theta)
\end{equation}
which makes the solution
\begin{equation}
A_{\mu}^a = \bar{\eta}_{\mu \nu}^a \, \Pi(x) \partial_\nu \Pi^{-1}(x) 
\end{equation} 
 time-independent,
where
\begin{equation}
\theta = \arctan\frac{\sin(\beta t_E)\sinh(\beta r)}{\cosh(\beta r) \cos(\beta t_E) - 1}
\end{equation}

We could try to perform the analogous procedure in our case for the S- Skyrmion constant in $x_3$.
The only problem is the change of the asymptotic behavior of our solution as $z \rightarrow \infty$ under the gauge transform (\ref{rossigauge}). We need to check that it is compatible with the approximation of the flat z coordinate near $z=0$, made in \cite{more}. The form of the gauge transformation (\ref{rossigauge}) will  be valid only for small values of $z$. We shall investigate this subtle
point elsewhere.
The energy density of such "monopole string" is finite:
\begin{equation}
\frac{dE}{dx_3} = \frac{dS}{dt_E dx_3} = \frac{4 \pi \beta}{g_{5d}^2}
\end{equation}

\subsection{Time crystal and S-branes} 

Recently the idea of the "`time crystal"' that is the dynamically
organized time periodicity has been forwarded \cite{wilchek} (see \cite{sasha} for review).
It was recognized that the time crystal is impossible in the equilibrium state
however the possibility for such state at non-equilibrium can not be excluded.
Moreover the scenario for the time crystal in the MBL state with periodic
quench has been suggested and observed experimentally. Such system develops
periodicity in time with period different from the period of the external
drive. 
Let us remark that the S-branes can serve as the building blocks in 
a kind of the time crystal in the constant external electric field in the
Euclidean space-time. 
Let us assume for example that the electric field is added to 1+1 theory with fermions and the
Schwinger pair creation is considered. The leading 
Euclidean bounce is just the circle \cite{alvarez} 
with negative mode 
however more general configuration involving the multiple 
parallel
extended instantons can be considered as well.
Such configuration has been discussed in \cite{array}.
To some extend one could say that the fermion-antifermion pair
interact instantly via the extended instanton which 
can be thought of as the bound state.

A bit loosely one could say that we have the array of the 
interacting Wilson loops in the 2d Euclidean space-time.
Due to the interaction between the Wilson loops 
we get a kind of neutral S-meson. Similarly we can get 
in higher dimensions the similar bounce in the external
magnetic field describing the monopole pair creation.
In this case we get the interacting t'Hooft loops and
the magnetic S-meson.

The emerging period in the time direction depends on the ratio
of the masses of particles and the tension of the S-meson. Upon summation
over the time ladder the emerging period looks like the
Unruch temperature for the accelerated particle in the
external electric field \cite{array}.
We get structure similar to the time crystal   in the Euclidean time but there is the
remnant of the dynamical period in the Minkowski time upon the
analytic continuation of the caterpillar bounce back to the Minkowski
time. Indeed the analytic continuation remembers the periodicity
in the Euclidean time and to some extend the Unruch temperature
is the counterpart of the time crystal period in the Euclidean space.
Note that our picture is different from the holographic picture for Floquet states
based on the Schwinger process in the time-periodic external electric field
which has been suggested in \cite{hashimoto}.

\section{Conclusion}

In this note we have found a new   space-like defect - S-Skyrmion
in the confined Euclidean QCD which can be thought of as the 
flavored extended instanton. The solution is localized in time
and hosts the several types of currents which are the counterparts
of the Skyrmion charges - baryonic, electric and axial. The 
baryonic current of the S-Skyrmion is quantized in the proper normalization 
and is topologically protected. If we use the dyonic instanton
solution in holographic QCD for S-Skyrmion the situation 
is quite close to the initial formulation of the position
of S- brane at the extremum of the tachyonic potential.

This topologically  quantized current has  similarities
with the Thouless current pump phenomena for the periodically driven systems
since the similar Chern number does the job. In the Thouless
case the electric charges emerge at the boundaries of the
gapped bulk. One more phenomena of the similar nature - the quantum
phase slip involves the "`creation"' of the vortex pair at the
edges of the gapped bulk via tunneling. We could speculate that
the S-Skyrmion process could produce the pair of magnetic monopoles
at the edges since we have shown that the S-Skyrmion can
terminate at the monopole via the dual version of the Callan-Rubakov effect.

Here we have considered the simplest example of 1d extended instanton
but the similar solutions involving the space-like defects with 
2d and 3d worldvolumes do exist as well. The lattice studies 
indicate that they exist on the equal footing with 1d defects and we hope to discuss
them elsewhere. 
We expect that the properties of the S-Skyrmion can be quite 
precisely analyzed in lattice QCD. Presumably the space-like
defects could also play the role in explanation of the well established
counterintuitive delocalization of all Dirac operator
modes in confined Euclidean QCD in a kind of fracton picture.

The role of these solutions
at non-vanishing temperature and chemical potential in particular
near the deconfinement phase transition should be clarified.  It would be also 
interesting to investigate carefully the structure
of the moduli spaces of such class of topological solutions
and the possibility of the network involving the 
defects of different codimensions to exist. Since the S-defects
generically carry the p-form current it is necessary to 
investigate carefully the current matching in the generic
network. Some work in this direction has been recently done 
in \cite{seiberg}. The S-brane can produce the conventional
time-like defects hence the possibility of more general
networks in the Euclidean QCD certainly has to be investigated.
The instantaneous S-defect could induce the interaction between
the higher dimensional branes like the instant induces the
t'Hooft vertex.

The moduli space of conventional Skyrmion
are quantized yielding the spectrum of excitations. Hence we 
expect that the similar quantization of the S-Skyrmion moduli space
has to be performed and will provide the tower of the excited
states of S-Skyrmion.

The chiral condensate is the analogue of the exciton condensate in the
solid state physics. Hence it would be interesting to elaborate the
similar S-Skyrmions in the condmat context. In particular the S-Skyrmions
presumably could yield a kind of the quantum phase slim phenomena in the excitonic condensate.

The work of A.Gor. was supported
by Basis Foundation fellowship and RFBR grant 19-02-00214. 
The work of A.Gr. was supported by the Russian
Science Foundation under grant 19-11-00062.
A.Gor. thanks Simons Center for Geometry and Physics at Stony Brook
University where the work has been completed for the hospitality and support.

\end{document}